\documentclass[11pt]{article}

\usepackage{amsmath,amssymb}
\usepackage{QED}
\usepackage{tikz}
\usetikzlibrary{automata}
\usepackage[utf8]{inputenc}

\newtheorem{definition}{Definition}

\newtheorem{lemma}[definition]{Lemma}
\newtheorem{theorem}[definition]{Theorem}

\newcommand{\pZ}{{\sf Player 0}}
\newcommand{\pO}{{\sf Player 1}}
\newcommand{\N}{\mathbb{N}}

\usepackage{anysize}
\marginsize{2.5cm}{2.5cm}{2.5cm}{2.5cm}

\author{St\'ephane Le Roux\\leroux@lsv.de}

\title{Memoryless Determinacy of Infinite Parity Games:\\Another Simple Proof}

\begin{document}
\maketitle

\begin{abstract}
In 1998 Zielonka simplified the proofs of memoryless determinacy of \emph{infinite} parity games. In 2018 Haddad simplified some proofs of memoryless determinacy of \emph{finite} parity games. This article adapts Haddad's technique for \emph{infinite} parity games. Two proofs are given, a shorter one and a more constructive one. None of them uses Zielonka's traps and attractors. %Furthermore in the shorter one, the induction on the number of priorities and the transfinite induction on the vertices are no longer interleaved. 
\end{abstract}

Keywords: positional determinacy; finitely many priorities; infinite graph; (transfinite) induction.

\section{Introduction}

The memoryless determinacy of infinite parity games with finitely many priorities was proved independently by Emerson and Jutla~\cite{EJ91}, and Mostowski~\cite{Mostowski91}, with various applications in computer science. Then Zielonka~\cite{Zielonka98} provided an elegant and simple argument for the same theorem. 

Several simpler proofs can be found in the literature in the case where the underlying graph is finite. These proofs usually proceed by induction on the number of relevant vertices. Recently, Haddad~\cite{Haddad18} found a very simple argument by defining accurately what a relevant vertex is: one that has incoming edges and proper outgoing edges. His proof splits one relevant vertex into two non-relevant ones for the induction step, and concludes after a case disjunction.

Since Haddad~\cite{Haddad18} also proceeds by induction on the number of relevant vertices, it cannot be used verbatim to prove the determinacy of infinite games with finitely many priorities. However, this article adapts Haddad's technique for these games by performing an induction on the number of relevant priorities, i.e. the priorities that label relevant vertices. For this purpose, Haddad's single vertex split is replaced with the splitting of (in)finitely many vertices at once.

Zielonka~\cite{Zielonka98} provided two proofs for the same theorem, a more constructive one and a shorter one, both highly relying on the notions of trap, attractor, etc. This article also provides a more constructive one and a shorter one. Both new proofs adapt Haddad's technique for infinite games, in two fairly different ways but without traps or attractors. Both more constructive proofs, in~\cite{Zielonka98} and in this article, nest a transfinite induction on the vertices within the induction on the finite number of priorities, and in turn the finite induction hypothesis is invoked within the transfinite induction. Unlike these, the shorter proof of this article avoids the second nesting and factors out a unique transfinite induction into Lemma~\ref{lem:max-win} on prefix-independent winning conditions. %(Note that Zielonka's shorter proof could also be adapted this way.)

The same theorem was generalised by Gr\"adel and Walukiewicz~\cite{GW06} for infinite sets of priorities, but the proofs from this article could not be used verbatim to prove this generalisation, since they proceed by induction on the number of relevant priorities.

Section~\ref{sect:tpg} includes general definitions on games and Lemmas~\ref{lem:max-win} and \ref{lem:less-self} about prefix-independent winning conditions. (These lemmas are probably folklore one way or another.) Section~\ref{sect:simple-proof} includes the definition of parity games and a shorter proof of memoryless determinacy of infinite parity games with finitely many priorities. Section~\ref{sect:more-constr} provides a more constructive proof.

\section{Two-player win/lose games}\label{sect:tpg}

A two-player win/lose game is a tuple  $\langle V_0,V_1,E,C,\pi, W\rangle$, where $V:= V_0 \sqcup V_1$ (disjoint union), and $E \subseteq V \times V$ satisfies $\forall v \in V,\exists u \in V,\,(v,u) \in E$, and $C \neq \emptyset$, and $\pi : V \to C$, and $W \subseteq C^\omega$.

{\bf The players} Informally, there are two players in such a game: \pZ\/ controlling the vertices in $V_0$, and \pO\/ controlling the vertices in $V_1$. 

{\bf Runs} A run in a game is an infinite path in the graph $(V,E)$, i.e. a sequence $\rho \in V^\omega$ such that $(\rho_{n}\rho_{n+1}) \in E$ for all $n \in \N$. The run $\rho$ is winning for \pZ\/ (\pO\/) if $\pi(\rho) \in W$ ($\notin W$), where $\pi(\rho) := \pi(\rho_0)\pi(\rho_1)\dots \in C^\omega$. Also, finite paths in $(V,E)$ are called finite runs.

{\bf Strategies} A \pZ\/ (\pO\/) memoryless strategy is a function $\sigma: V_0 \to V$ ($\tau: V_1 \to V$) such that $(v,\sigma(v)) \in E$ ($(v,\tau(v)) \in E$) for all $v \in V_0$ ($V_1$). Only memoryless strategies are considered in this article, where they are often simply called strategies. 

{\bf Compatible runs} A run $\rho$ is compatible with a strategy $\sigma: V_0 \to V$ ($\tau: V_1 \to V$) if $\rho_{n+1} = \sigma(\rho_n)$ whenever $\rho_n \in V_0$ ($V_1$). This notion is naturally extended to finite runs.

{\bf Winning strategies and regions} A \pZ\/ strategy $\sigma$ is said to win from some $v \in V$ if all compatible runs $\rho$ such that $\rho_0 = v$ are winning for \pZ\/. Let $W^{m}_0(G,\sigma)$ be the vertices from where $\sigma$ wins, and let $W^{m}_0(G)$ be the vertices from where some \pZ\/ strategy wins. The $m$ in $W^{m}$ stands for memoryless. A \pZ\/ strategy $\sigma$ is said to be memoryless optimal (optimal for short) if $W^{m}_0(G,\sigma) = W^{m}_0(G)$. Likewise $W^{m}_1(G,\tau)$ and  $W^{m}_1(G)$ are defined for \pO\/. (It is straightforward to prove that $W^{m}_0(G) \cap W^{m}_1(G) = \emptyset$.)

%{\bf Suffix-closed and prefix-independent winning condition} $W$ is called suffix-closed if removing finite prefixes from runs preserves membership to $W$, i.e. $\forall \rho \in C^\omega, \forall w \in C^*,\, w\rho \in W \,\Rightarrow\, \rho \in W$. It is called prefix-independent if both $W$ and $C^\omega \setminus W$ are suffix-closed.
{\bf Prefix-independent winning condition} $W$ is called prefix-independent if removing or adding prefixes to runs preserves membership to $W$, i.e. $\forall (\rho,w) \in C^\omega\times C^*,\, w\rho \in W \,\Leftrightarrow\, \rho \in W$.

Lemma~\ref{lem:max-win} below is only used in Section~\ref{sect:simple-proof}. It states that for all games with prefix-independent winning condition, there exists a memoryless strategy that wins from all vertices from where the player can win without memory. It shares similarities with \cite[Second proof, p150]{Zielonka98}.
\begin{lemma}\label{lem:max-win}
Let $W \subseteq C^\omega$ be a prefix-independent winning condition. For all games $G$ using colors $C$ and winning condition $W$, there exists a \pZ\/ memoryless strategy $\sigma$ such that $W^{m}_0(G,\sigma) = W^{m}_0(G)$. (And likewise for \pO\/.)
\begin{proof}
Let $\alpha$ be some ordinal of cardinality $|W^{m}_0(G)|$, and let $\{v_\beta\}_{\beta \in \alpha}$ be an enumeration of $W^{m}_0(G)$.  For all $\beta \in \alpha$ let $\sigma_\beta$ be a \pZ\/ strategy that is winning from $v_\beta$. Let $f: W^{m}_0(G) \to \alpha$ be such that $f(v)$ is the least $\beta$ with $v \in W^{m}_0(G,\sigma_\beta)$, which is well-defined since every non-empty subset of an ordinal has a least element. Let $\sigma$ be a \pZ\/ strategy satisfying $\sigma(v) := \sigma_{f(v)}(v)$ for all $v \in V_0 \cap W^{m}_0(G)$.

Let a run $\rho$ start in $W^{m}_0(G)$ and be compatible with $\sigma$. If $\rho$ is also compatible with $\sigma_{f(\rho_0)}$, it makes \pZ\/ win, by definition of $f$. Otherwise let $k$ be the least such that $\rho_k \in V_0$ and $\rho_{k+1} \neq \sigma_{f(\rho_0)}(\rho_k)$, so $f(\rho_k) < f(\rho_0)$ by definition of $\sigma$. To apply the argument recursively, note that $\rho_k \in W^{m}_0(G,\sigma_{f(\rho_0)}) \subseteq W^{m}_0(G)$ since $\rho_k$ is compatible with $\sigma_{f(\rho_0)}$ and by prefix independence. By property of the ordinals this situation can occur only finitely many times, so the tail of $\rho$ is eventually compatible with some $\sigma_{f(\rho_n)}$ from some $\rho_n$ on. So by prefix-independence, $\rho$ makes \pZ\/ win.
\end{proof}
\end{lemma}

{\bf Uniform memoryless determinacy} A game $G = \langle V_0,V_1,E,C,\pi, W\rangle$ is uniformly memoryless determined if there exist \pZ\/ and \pO\/ memoryless strategies $\sigma$ and $\tau$, respectively, such that $W^m_0(G,\sigma) \cup W^m_1(G,\tau) = V_0 \cup V_1$.

The remainder of Section~\ref{sect:tpg} is only used in Section~\ref{sect:more-constr}.

{\bf Unfair-win vertex} For all games $\langle V_0,V_1,E,C,\pi, W\rangle$ a vertex $v \in V_0$ (resp. $V_1$) is called an unfair win if it has a self-loop and proper outgoing edges, and $\pi(v)^\omega \in W$ (resp. $C^\omega \setminus W$).

Lemma~\ref{lem:less-self} below states that when trying to prove memoryless determinacy of prefix-independent winning conditions, it suffices to consider games void of unfair-win vertices.
\begin{lemma}\label{lem:less-self}
%Fix $C$ and $W \subseteq C^\omega$ such that $C^\omega \setminus W$ is suffix-closed. If for all games $\langle V_0,V_1,E,C,\pi, W\rangle$ that are void of unfair-win vertices there exist \pZ\/ and \pO\/ memoryless strategies $\sigma$ and $\tau$, respectively, such that $W^m_0(G,\sigma) \cup W^m_1(G,\tau) = V_0 \cup V_1$, it also holds without the unfair-win restriction.
Fix $C$ and a prefix-independent $W \subseteq C^\omega$. If the games $G = \langle V_0,V_1,E,C,\pi, W\rangle$ void of unfair-win vertices are uniformly memoryless determined, so are all games, i.e. without the unfair-win restriction.
\begin{proof}
Let us transform an arbitrary game $G = \langle V_0,V_1,E,C,\pi,W\rangle$ into $G^- =\langle V_0,V_1,E^-,C,\pi,W\rangle $ by removing the proper outgoing edges of the unfair-win vertices, thus making them absorbing, as in Figure~\ref{fig:g2g-}. % For all $i \in \{0,1\}$, for all $v \in V_i$ that has a self-loop, if the priority on $v$ makes the controlling player win (i.e. $\pi(v) = i \mod 2$), let us make $v$ an absorbing vertex (i.e. $vE^- := \{v\}$), as in Figure~\ref{fig:g2g-}.% If the priority on $v$ makes the non-controlling player win (i.e. $\pi(v) = 1-i \mod 2$), let us remove the self-loop on $v$ (i.e. $vE^- := vE \setminus\{v\}$), as in Figure~\ref{fig:g2g-} again.
Let $\sigma$ be a \pZ\/  memoryless strategy in $G^-$, so $\sigma$ is also a strategy in $G$. Let a run $\rho$ start in $W^m_0(G^-,\sigma)$ and be compatible with $\sigma$ in $G$. Let us show that $\rho$ is also compatible with $\sigma$ in $G^-$. Towards a contradiction, let $(\rho_n,\rho_{n+1})$ be the first edge of $\rho$ that is not present in $E^-$, so $\rho_n$ is a unfair-win vertex in $G$, and $\rho_n \in V_1$ (as \pZ\/ just follows $\sigma$), so $\pi(\rho_n)$ is odd. It implies that the prefix $\rho_{\leq n}$, which is compatible with $\sigma$ in $G^-$ by choice of $n$, leads in $G^-$ to an absorbing state with odd priority, contradicting (by prefix independence) the assumption that $\rho_0 \in W^m_0(G^-,\sigma)$. Therefore $\rho$ is also compatible with $\sigma$ in $G^-$ and thus makes \pZ\/ win, which shows that $W^m_0(G^-,\sigma) \subseteq W^m_0(G,\sigma)$ for all $\sigma$. By symmetry $W^m_1(G^-,\tau) \subseteq W^m_1(G,\tau)$ for all \pO\/ strategy $\tau$. By assumption let $\sigma$ and $\tau$ be \pZ\/ and \pO\/ memoryless strategies, respectively, such that $W^m_0(G^-,\sigma) \cup W^m_1(G^-,\tau) = V_0 \cup V_1$. So $W^m_0(G,\sigma) \cup W^m_1(G,\tau) = V_0 \cup V_1$ by the above inclusions. 
\begin{figure}
\centering
\begin{minipage}{1.8in}
\begin{tikzpicture}[shorten >=1pt,node distance=1.8cm, auto]
    \node[state, circle split, label = below:$v$] (m) {i\nodepart{lower} $\pi(v)$};
  \node[state, draw = none] (tl) [above left of = m]{};
  \node[state, draw = none] (bl) [below left of = m] {};
    \node[state, draw = none] (tr) [above right of = m]{};
  \node[state, draw = none] (br) [below right of = m] {};
    
\path[->] (tl) edge[dashed] node {} (m)		
		(bl) edge[dashed] node {} (m)
		(m) edge node {} (tr)		
		(m) edge[dashed] node {} (br);
\draw[->] (m) edge [loop above] node{} ();
 \end{tikzpicture}
 \end{minipage}
 \begin{minipage}{1.8in}
\begin{tikzpicture}[shorten >=1pt,node distance=1.8cm, auto]
    \node[state, circle split, label = below:$v$] (m) {i\nodepart{lower} $\pi(v)$};
  \node[state, draw = none] (tl) [above left of = m]{};
  \node[state, draw = none] (bl) [below left of = m] {};
    \node[state, draw = none] (tr) [above right of = m]{};
  \node[state, draw = none] (br) [below right of = m] {};
    
\path[->] (tl) edge[dashed] node {} (m)		
		(bl) edge[dashed] node {} (m);
\draw[->] (m) edge [loop above] node{} ();
 \end{tikzpicture}
 \end{minipage}
 \begin{minipage}{1.8in}
\begin{tikzpicture}[shorten >=1pt,node distance=2cm, auto]
    \node[state, circle split, label = below:$v$] (m) {i\nodepart{lower} $\pi(v)$};
  \node[state, draw = none] (tl) [above left of = m]{};
  \node[state, draw = none] (bl) [below left of = m] {};
    \node[state, draw = none] (tr) [above right of = m]{};
  \node[state, draw = none] (br) [below right of = m] {};
    
\path[->] (tl) edge[dashed] node {} (m)		
		(bl) edge[dashed] node {} (m)
		(m) edge node {} (tr)		
		(m) edge[dashed] node {} (br);
 \end{tikzpicture}
 \end{minipage}
\caption{From $G$ (left-hand side), removing edges from unfair-win vertices if $\pi(v)^\omega$ makes Player $i$ win (middle) or useless self-loops if $\pi(v) \neq i \mod 2$ (right-hand side).}
\label{fig:g2g-}
 \end{figure}
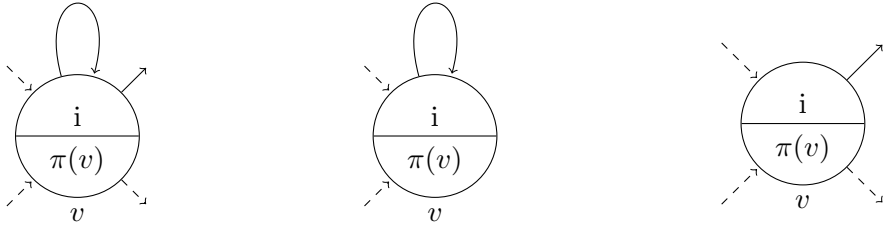
 \end{proof}
\end{lemma}

\section{A shorter proof}\label{sect:simple-proof}

{\bf Parity games} A parity game with finitely many priorities is a two-player win/lose game where $C = \N$, the function $\pi : V \to \N$ is bounded, and the prefix-independent $W$ is defined as follows. For all runs $\rho$ let $\max_{\infty}(\rho) := \max\{k \in \N \mid \forall i \in \N,\exists j > i,\pi(\rho_j) = k \}$, which is well-defined since $\pi$ is bounded. The run $\rho$ is winning for \pZ\/ (\pO\/) if $\max_{\infty}(\rho)$ is even (odd). So a parity game amounts to a tuple $\langle V_0,V_1,E,\pi\rangle$ where $\pi : V \to \N$ is bounded.

{\bf Absorbing, vanishing, and relevant vertices \cite{Haddad18}} A vertex $v \in V$ is absorbing if it has no proper outgoing edge, i.e. $\{w \mid (v,w) \in E\} = \{v\}$; it is vanishing if $\{w \mid (w,v) \in E\} = \emptyset$, i.e. it has no incoming edge; it is relevant if it is neither absorbing nor vanishing.

{\bf Relevant priorities} Let $V_r$ be the set of the relevant vertices of a parity game. Then $\{\pi(v)\mid v \in V_r\}$ is the set of the relevant priorities.

Theorem~\ref{thm:short-proof} below is proved without interleaving the inductions.

\begin{theorem}[\cite{Zielonka98}]\label{thm:short-proof}
The parity games are uniformly memoryless determined.
\begin{proof}
Let us prove the claim by induction on the number of relevant priorities. Base case: if there are no relevant priorities in a game, there is no relevant vertices either, so all vertices are either absorbing or vanishing. In this case every run visits at most two vertices, and it is straightforward to prove the claim by backward induction.

For the inductive case, let us assume that the claim holds for all games with relevant priorities less than some $k$, and let us proceed in two steps. The first step will show that $W^{m}_0(G) \cup W^{m}_1(G) \neq \emptyset$ for all games $G = \langle V_0,V_1,E,\pi\rangle$ with relevant priorities at most $k$, and the second step will invoke Lemma~\ref{lem:max-win} to complete the proof.

{\bf First step} Let $D$ be the relevant vertices from $V$ with priority $k$. Wlog let us assume that $k$ is even, i.e. up to adding $1$ to every priority and swapping $V_0$ and $V_1$.

Let us modify $G$ to make the priority $k$ irrelevant and to be able to invoke the induction hypothesis. More specifically, let us derive $G^+$ from $G$ by splitting each vertex $v \in D$ into one vanishing split vertex $v$ that keeps the proper outgoing edges and the same controller, and one absorbing split vertex $\tilde{v} \notin V$ that keeps the incoming edges and receives a self-loop. (The controller of $\tilde{v}$ and the priority of $v$ are irrelevant, so, e.g., both split vertices preserve them.) This is depicted in Figure~\ref{fig:g2g+}.
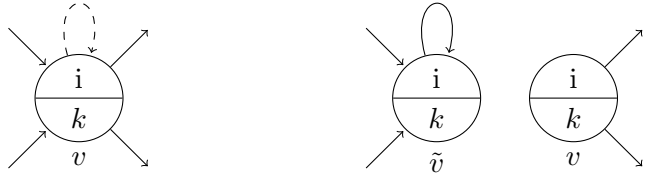
\begin{figure}
\centering
\begin{minipage}{1.8in}
\begin{tikzpicture}[shorten >=1pt,node distance=1.8cm, auto]
    \node[state, circle split, label = below:$v$] (m) {i\nodepart{lower} $k$};
  \node[state, draw = none] (tl) [above left of = m]{};
  \node[state, draw = none] (bl) [below left of = m] {};
    \node[state, draw = none] (tr) [above right of = m]{};
  \node[state, draw = none] (br) [below right of = m] {};
    
\path[->] (tl) edge node {} (m)		
		(bl) edge node {} (m)
		(m) edge [loop above, dashed] node{} ()
		(m) edge node {} (tr)		
		(m) edge node {} (br);
%\draw[->, dashed] (m) edge [loop above] node{} ();
 \end{tikzpicture}
 \end{minipage}
\begin{minipage}{1.5in}
\begin{tikzpicture}[shorten >=1pt,node distance=1.8cm, auto]
    \node[state, circle split, label = below:$\tilde{v}$] (m) {i\nodepart{lower} $k$};
      \node[state, circle split, label = below:$v$] (m2) [right of = m]{i\nodepart{lower} $k$};
  \node[state, draw = none] (tl) [above left of = m]{};
  \node[state, draw = none] (bl) [below left of = m] {};
    \node[state, draw = none] (tr) [above right of = m2]{};
  \node[state, draw = none] (br) [below right of = m2] {};
    
\path[->] (tl) edge node {} (m)		
		(bl) edge node {} (m)
		(m) edge [loop above] node{} ()
		(m2) edge node {} (tr)		
		(m2) edge node {} (br);
 \end{tikzpicture}
\end{minipage}
\caption{Splitting a vertex $v$ controlled by Player $i$ and with priority $k$: from $G$ (left-hand side) to $G^+$ (right-hand side). The dashed arrow means that a self-loop may or may not be present.}
\label{fig:g2g+}
 \end{figure}
Formally, for all $i \in \{0,1\}$ let  $V^+_i := V_i \cup \{\tilde{v}\,\mid\,v \in D \cap V_i\}$, where $\tilde{D} := \{\tilde{v}\,\mid\,v \in D\}$ and $V \cap \tilde{D} = \emptyset$. Also let $V^+ := V^+_0 \cup V^+_1$ and $E^+ := \{(u,v) \in E\mid v \notin D\} \cup \{(u,\tilde{v})\mid (u,v) \in E \wedge v \in D\} \cup \tilde{D}^2$. Let $\pi^+: V^+ \to \N$ be defined by $\pi^+\mid_{V} := \pi$ and $\pi^+(\tilde{v}) := k$ for all $v \in D$, and let $G^+ :=  \langle V^+_0,V^+_1,E^+,\pi^+\rangle$. 

The new game $G^+$ has one less relevant priority than $G$, namely $k$, so by induction hypothesis there exist \pZ\/ and \pO\/ memoryless strategies $\sigma^+$ and $\tau^+$, respectively, such that $W^{m}_0(G^+,\sigma^+) \cup W^{m}_1(G^+,\tau^+) = V^+_0 \cup V^+_1$.  Either $\sigma^+$ or $\tau^+$ will induce a winning strategy in $G$, which will complete the first step.

Before letting the players use $\sigma^+$ or $\tau^+$ and play in $G$, the strategies need modifying, as their domains and especially codomains are different in $G$ and $G^+$ (i.e. $V$ vs $V^+ = V \cup \tilde{D}$). To prepare the modification, let $f: V \cup \tilde{D} \to V$ be defined by $f(v) := v$ for all $v \in V$, and $f(\tilde{v}) : = v$ for all $v \in D$. Now, the function $f \circ \sigma^+\mid_{V_0}^{V}: V_0 \to V$ is a \pZ\/ strategy in $G$, and $f \circ \tau^+\mid_{V_1}^{V}: V_1 \to V$ is a \pO\/ strategy in $G$. The effect of $f$ is depicted in Figure~\ref{fig:s+2s}, where the double lines represent partial strategies towards $\tilde{D}$ and $D$, respectively.
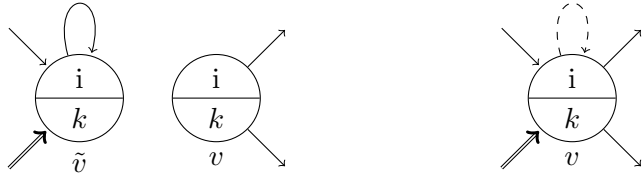
\begin{figure}
\centering
\begin{minipage}{2.5in}
\begin{tikzpicture}[shorten >=1pt,node distance=1.8cm, auto]
    \node[state, circle split, label = below:$\tilde{v}$] (m) {i\nodepart{lower} $k$};
      \node[state, circle split, label = below:$v$] (m2) [right of = m]{i\nodepart{lower} $k$};
  \node[state, draw = none] (tl) [above left of = m]{};
  \node[state, draw = none] (bl) [below left of = m] {};
    \node[state, draw = none] (tr) [above right of = m2]{};
  \node[state, draw = none] (br) [below right of = m2] {};
    
\path[->] (tl) edge node {} (m)		
		(bl) edge [double] node {} (m)
		(m) edge [loop above] node{} ()
		(m2) edge node {} (tr)		
		(m2) edge node {} (br);
 \end{tikzpicture}
\end{minipage}
\begin{minipage}{1.5in}
\begin{tikzpicture}[shorten >=1pt,node distance=1.8cm, auto]
    \node[state, circle split, label = below:$v$] (m) {i\nodepart{lower} $k$};
  \node[state, draw = none] (tl) [above left of = m]{};
  \node[state, draw = none] (bl) [below left of = m] {};
    \node[state, draw = none] (tr) [above right of = m]{};
  \node[state, draw = none] (br) [below right of = m] {};
    
\path[->] (tl) edge node {} (m)		
		(bl) edge[double]  node {} (m)
		(m) edge [loop above, dashed] node{} ()
		(m) edge node {} (tr)		
		(m) edge node {} (br);
%\draw[->, dashed] (m) edge [loop above] node{} ();
 \end{tikzpicture}
 \end{minipage}
\caption{Merging back vertices, from $\sigma^+$ (left-hand side) to $f \circ \sigma$ (right-hand side)}
\label{fig:s+2s}
 \end{figure}
  
An important property of $f$ is that a (finite) run $\rho \in V((V\setminus D)^* \cup (V\setminus D)^\omega)$ is compatible with $\sigma^+$ (resp. $\tau^+$) in $G^+$ iff it is compatible with $f \circ \sigma^+$ (resp. $f \circ \tau^+$) in $G$. Indeed,
\begin{align}
\rho \mbox{ is compatible with } \sigma^+ & \quad\Leftrightarrow\quad \forall n, \, \rho_n \in V_0\,\Rightarrow\, \rho_{n+1} = \sigma^+(\rho_n) \mbox{ by definition,} \nonumber\\
	&  \quad\Leftrightarrow\quad \forall n, \, \rho_n \in V_0\,\Rightarrow\, \rho_{n+1} = f \circ \sigma^+(\rho_n) \mbox{ since } \rho_{n+1} \in V \setminus D,\nonumber\\
	&  \quad\Leftrightarrow\quad \rho \mbox{ is compatible with } f \circ \sigma^+. \label{eq:comp-run}
\end{align}

To prove the forthcoming Inequality~(\ref{eq:non-empty}) let us make a case disjunction. First case, $W^{m}_0(G^+,\sigma^+) = V^+$. Let $\rho$ be a run compatible with $f \circ \sigma^+$ in $G$. If $\rho$ sees priority $k$ infinitely often, it makes \pZ\/ win. If $\rho$ sees priority $k$ finitely many times only, $\rho$ visits $D$ only finitely often, so from some point on, $\rho$ is compatible with $\sigma^+$, by Equivalence~(\ref{eq:comp-run}). By prefix-independence and since $\sigma^+$ is winning from everywhere, $\rho$ makes \pZ\/ win. This shows that $f \circ \sigma^+$ wins in $G$ (from everywhere), thus implying Inequality~(\ref{eq:non-empty}).

Second case,  $W^{m}_0(G^+,\sigma^+) \neq V^+$, so let $v \in W^{m}_1(G^+,\tau^+) = V^+ \setminus W^{m}_0(G^+,\sigma^+)$, and let $\rho$ be a run that starts at $v$ and that is compatible with $f \circ \tau^+$ in $G$.  Towards a contradiction let us assume that $\rho$ enters $D$ for the first time at time $n+1$. It implies that the prefix $\rho_{\leq n}$ is compatible with $\tau^+$ by Equivalence~(\ref{eq:comp-run}), and that $\tau^+(\rho_n) = \tilde{\rho}_{n+1}$. Subsequently, it implies that $\rho_{\leq n}(\tilde{\rho}_{n+1})^\omega$ is compatible with $\tau^+$, whereas it makes \pZ\/ win, contradiction, so $\rho$ does not enter $D$. So $\rho$ is compatible with $\tau^+$ by Equivalence~(\ref{eq:comp-run}), thus making \pO\/ win. This shows that $f \circ \tau^+$ wins in $G$ from $v$, thus implying Inequality~(\ref{eq:non-empty}).
Therefore, for all games $G$ with relevant priorities at most $k$,
\begin{align}
W^{m}_0(G) \cup W^{m}_1(G) \neq \emptyset\label{eq:non-empty}
\end{align}

{\bf Second step} Let $G$ be a game with relevant priorities at most $k$. By Lemma~\ref{lem:max-win} let $\sigma$ and $\tau$ be \pZ\/ and \pO\/ memoryless strategies, respectively, such that $W^{m}_0(G,\sigma) = W^{m}_0(G)$  and $W^{m}_1(G,\tau) = W^{m}_1(\tau)$. Towards a contradiction, which will prove the claim, let us assume that $W^{m}_0(G,\sigma) \cup W^{m}_1(G,\tau) \neq V_0 \cup V_1$. Let $V^- := V \setminus W^{m}_0(G) \cup W^{m}_1(G)$ and let us argue two useful facts about $V^-$ (thus using one Zielonka's trap without defining the concept).
\begin{enumerate}
\item\label{fact:v-1} For all $v \in V^-$ there exists $u \in V^-$ such that $(v,u) \in E$. Otherwise the player controlling $v$ would have no other choice than entering $W^{m}_0(G) \cup W^{m}_1(G)$, which would imply $v \in W^{m}_0(G) \cup W^{m}_1(G)$, a contradiction.

\item\label{fact:v-2}  If $v \in V_i \cap V^-$, there is in $G$ no edge from $v$ to $ W^{m}_i(G)$. Otherwise Player $i$ could reach $W^{m}_i(G)$ from $v$, thus implying $v \in W^{m}_i(G)$, a contradiction since $V^- \cap W^{m}_i(G) = \emptyset$.  
\end{enumerate}
Let the game $G^-$ be the restriction of $G$ to $V^-$, i.e. $G^-:= \langle V_0^-,V_1^-,E^-,\pi^-\rangle$, where $V_0^- := V_0 \cap V^-$ and $V_1^- := V_1 \cap V^-$ and $E^- := E \cap(V^- \times V^-)$, and $\pi^- := \pi\mid_{V^-}$. By Fact~\ref{fact:v-1} above, $G^-$ is indeed a parity game. By Fact~\ref{fact:v-2} above, $W^{m}_i(G^-) \subseteq W^{m}_i(G)$ for all $i$, so  by definition of $V^-$
\begin{align}
W^{m}_0(G^-) \cup W^{m}_1(G^-) = \emptyset\label{eq:empty}
\end{align}
However, the game $G^-$ uses priorities at most $k$, so Inequality~(\ref{eq:non-empty}) contradicts Equality~(\ref{eq:empty}) and the assumption $W^{m}_0(G,\sigma) \cup W^{m}_1(G,\tau) \neq V_0 \cup V_1$ made at the beginning of the second step.
\end{proof}
\end{theorem}

Note that the first step of the proof of Theorem~\ref{thm:short-proof}, which uses Haddad's technique, could be replaced with something similar to the end of \cite[First proof, p149]{Zielonka98} from Equation $(5)$ onwards.

%\begin{proof}
%Let us prove differently the first step of the proof of Theorem~\ref{thm:short-proof}. Wlog let us assume that $k$ is even, i.e. up to adding $1$ to every priority and swapping $V_0$ and $V_1$. 

%Let $K := \pi^{-1}(k)$ and let $K^{+}$ be the set of vertices from which \pO\/ can force the play to reach $K$, which she can do via a memoryless strategy $\sigma^K$ with domain $K^+$. Let $V^- := V \setminus K^{+}$.

%Let $G^- := \langle V_0^-,V_1^-,E^-,\pi^-\rangle$, where blabla.

%By induction hypothesis let $\sigma^-$ and $\tau^-$ be \pZ\/ and \pO\/ memoryless strategies in $G^-$ such that $W^m_0(G^-,\sigma^-) \cup W^m_1(G^-,\tau^-) = V_0^- \cup V_1^-$. Let us make a case disjunction. First case, $W^m_0(G^-,\sigma^-) = V^-$, so $W^m_0(G,\sigma^- \cup \sigma^K) = V$, which shows (more than) the claim. 

%Second case, there exists $v \in V^-$ such that $v \in W_1^m(G^-, \tau^-)$ . Let $\tau$ be a \pO\/ memoryless strategy that extends $\tau^-$. Thus $v \in W^m_1(G,\tau)$, which shows the claim.
%\end{proof}

\section{A more constructive proof}\label{sect:more-constr}

{\bf Useless self-loops} For all parity games $\langle V_0,V_1,E,\pi\rangle$ an edge $(v,v)$ with $v \in V_0$ (resp. $V_1$) is called useless if proper outgoing edges also start from $v$, and $\pi(v)$ is odd (resp. even).

\begin{lemma}\label{lem:no-self-loop}
If the parity games where only absorbing states have self-loops are uniformly memoryless determined, so are all parity games.
\begin{proof}
By Lemma~\ref{lem:less-self} it suffices to show that the parity games void of unfair-win vertices are uniformly memoryless determined, since the parity condition is prefix independent. Note that a self-loop on a non-absorbing vertex is either useless or it makes the vertex unfair-win. 
 
Let us first transform an arbitrary parity game $G = \langle V_0,V_1,E,\pi\rangle$ into $G^- =\langle V_0,V_1,E^-,\pi\rangle$ by removing the useless self-loops, as in Figure~\ref{fig:g2g-} (righthand side). Let $\sigma$ be a \pZ\/  memoryless strategy in $G^-$, so $\sigma$ is also a strategy in $G$. Let $\rho$ be a run starting in $W^m_0(G^-,\sigma)$ and that is compatible with $\sigma$ in $G$, which is less constrained than being compatible with $\sigma$ in $G^-$: the run $\rho$ amounts to the interleaving of some run $\rho^-$ that is compatible with $\sigma$ on $G^-$, which makes \pZ\/ win, with finitely or infinitely many uses of useless self-loops on vertices controlled by \pO\/. These vertices have even priorities, so $\rho$ makes \pZ\/ win just like $\rho^-$ does, which shows that $W^m_0(G^-,\sigma) \subseteq W^m_0(G,\sigma)$ for all $\sigma$. By symmetry $W^m_1(G^-,\tau) \subseteq W^m_1(G,\tau)$ for all \pO\/ strategy $\tau$.

Let us now assume that $G$ is void of unfair-win vertices. So in $G^-$  only absorbing states have self-loops, so $G^-$ is uniformly memoryless determined by assumption. Let $\sigma$ and $\tau$ be \pZ\/ and \pO\/ memoryless strategies, respectively, such that $W^m_0(G^-,\sigma) \cup W^m_1(G^-,\tau) = V_0 \cup V_1$. So $W^m_0(G,\sigma) \cup W^m_1(G,\tau) = V_0 \cup V_1$ by the above inclusions.
\end{proof}
\end{lemma}

\begin{theorem}[\cite{Zielonka98}]\label{thm:long-proof}
The parity games are uniformly memoryless determined.
\begin{proof}
By Lemma~\ref{lem:no-self-loop} it suffices to prove the claim for games where only absorbing states have self-loops. Let us proceed by induction on the number of relevant priorities in $G$. The beginning of the argument is the same as the proof of Theorem~\ref{thm:short-proof} until Equivalence~\ref{eq:comp-run}. (The only difference is that there should not be dashed self-loops in the new Figures~\ref{fig:g2g+} and \ref{fig:s+2s}.) 

Unfortunately, composing the optimal strategies $\sigma^+$ and $\tau^+$ from $G^+$ with $f$ does not always yield optimal strategies for $G$. Figure~\ref{fig:problem-g-} exemplifies this with $k = 4$: going from $u$ to $\tilde{v}$, as suggested by the double arrow, is a \pZ\/ winning strategy in $G^+$ but going from $u$ to $v$ is not winning in $G$. %Indeed, starting from $v$, \pO\/ can already win in $G^+$. %This is especially problematic since \pO\/ can actually win from $u$ in $G$ by going to the left.
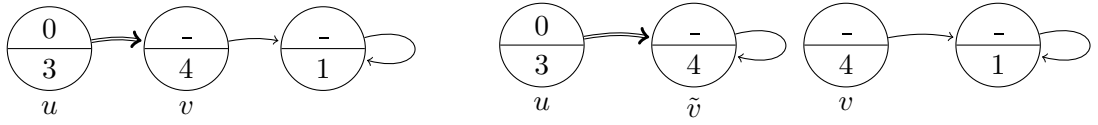
\begin{figure}
\centering
%\begin{tabular}{c}
\begin{minipage}{2.5in}
 \begin{tikzpicture}[shorten >=1pt,node distance=1.8cm, auto]
% \node[state, circle split] (ll){\_\nodepart{lower}2};
  \node[state, circle split] (lm) [label = below:$u$]{0\nodepart{lower}3};
%   \node[state, circle split] (lr) [right of = lm]{\nodepart{lower}4};
  %   \node[right of = lr](lrr){};
     \node[state, circle split, label = below:$v$] (rl) [right of = lm]{\_\nodepart{lower}4};
     \node[state, circle split] (rr) [right of = rl]{\_\nodepart{lower}1};
        
\path[->]	%(ll) edge [loop left] node{} () 	
		%(lm) edge [bend left = 10] node {} (ll)
		(lm) edge [double,bend left = 10] node{} (rl)
		(rl) edge [bend left = 10] node {} (rr)		
		(rr) edge [loop right] node{} ();
 \end{tikzpicture}
\end{minipage}
\begin{minipage}{2.5in}
\begin{tikzpicture}[shorten >=1pt,node distance=2cm, auto]
% \node[state, circle split] (ll){\_\nodepart{lower}2};
  \node[state, circle split] (lm) [label = below:$u$]{0\nodepart{lower}3};
   \node[state, circle split, label = below:$\tilde{v}$] (lr) [right of = lm]{\_\nodepart{lower}4};
  %   \node[right of = lr](lrr){};
     \node[state, circle split, label = below:$v$] (rl) [right of = lr]{\_\nodepart{lower}4};
     \node[state, circle split] (rr) [right of = rl]{\_\nodepart{lower}1};
        
\path[->]	%(ll) edge [loop left] node{} () 	
		%(lm) edge [bend left = 10] node {} (ll)
		(lm) edge [double,bend left = 10] node{} (lr)
		(lr) edge [loop right] node{} () 
		(rl) edge [bend left = 10] node {} (rr)		
		(rr) edge [loop right] node{} ();
 \end{tikzpicture}
 \end{minipage}
%  \end{tabular}
\caption{$G^+$ (right-hand side) is derived from $G$ (left-hand side).}
\label{fig:problem-g-}
 \end{figure}

To solve the above issue, $G^+$ is further modified as shown in Figure~\ref{fig:prob-sol}, where the parity of $i$ is irrelevant.
\begin{figure}
\centering
\begin{minipage}{2.5in}
\begin{tikzpicture}[shorten >=1pt,node distance=1.8cm, auto]
    \node[state, circle split, label = below:$\tilde{v}$] (m) {i\nodepart{lower} $k$};
      \node[state, circle split, label = below:$v$] (m2) [right of = m]{i\nodepart{lower} $k$};
  \node[state, draw = none] (tl) [above left of = m]{};
  \node[state, draw = none] (bl) [below left of = m] {};
    \node[state, draw = none] (tr) [above right of = m2]{};
  \node[state, draw = none] (br) [below right of = m2] {};
    
\path[->] (tl) edge node {} (m)		
		(bl) edge [double] node {} (m)
		(m) edge [loop above] node{} ()
		(m2) edge node {} (tr)		
		(m2) edge node {} (br);
 \end{tikzpicture}
\end{minipage}
\begin{minipage}{2.5in}
\begin{tikzpicture}[shorten >=1pt,node distance=1.9cm, auto]
    \node[state, circle split, label = below:$\tilde{v}$] (m) {i\nodepart{lower} $k+1$};
      \node[state, circle split, label = below:$v$] (m2) [right of = m]{i\nodepart{lower} $k$};
  \node[state, draw = none] (tl) [above left of = m]{};
  \node[state, draw = none] (bl) [below left of = m] {};
    \node[state, draw = none] (tr) [above right of = m2]{};
  \node[state, draw = none] (br) [below right of = m2] {};
    
\path[->] (tl) edge node {} (m)		
		(bl) edge [double] node {} (m)
		(m) edge [loop above] node{} ()
		(m2) edge node {} (tr)		
		(m2) edge node {} (br);
 \end{tikzpicture}
\end{minipage}
\caption{$G^+$ with $v \notin W^m_{(k \mod 2)}(G^+)$ (left-hand side) to $G_1$ (right-hand side)}
\label{fig:prob-sol}
 \end{figure}
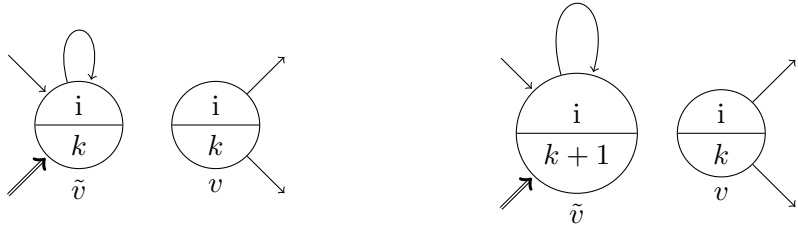
Yet, changing some priorities $k$ to $k+1$ may create new issues elsewhere, which in turn require to change more priorities $k$ to $k+1$. Transfinitely many modifications may even be required, as suggested in Figure~\ref{fig:transfinite}. Note that only the priorities of $G^+$ may be modified: its structure remains the same, so runs and strategies that are valid in one of the modifications are also valid in the other ones.
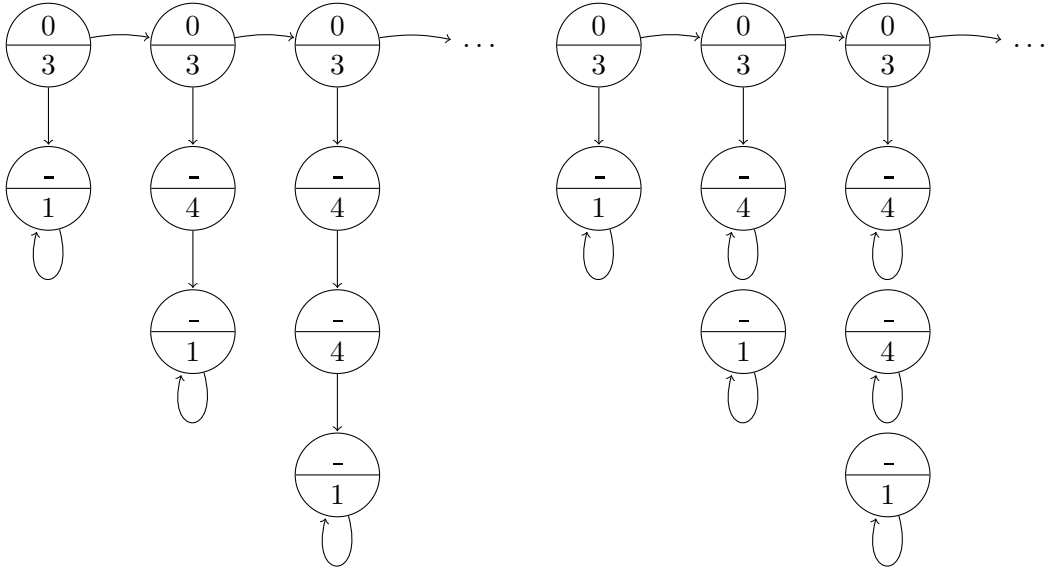
\begin{figure}
\centering
\begin{minipage}{2.8in}
 \begin{tikzpicture}[shorten >=.5pt,node distance=1.9cm, auto]
 \node[state, circle split] (n00){0\nodepart{lower}3};
 \node[state, circle split] (n01) [right of = n00]{0\nodepart{lower}3};
 \node[state, circle split] (n02) [right of = n01]{0\nodepart{lower}3};
 \node (n03) [right of = n02]{$\dots$};   
 \node[state, circle split] (n10) [below of = n00]{\_\nodepart{lower}1};  
 \node[state, circle split] (n11) [below of = n01]{\_\nodepart{lower}4};  
 \node[state, circle split] (n21) [below of = n11]{\_\nodepart{lower}1};  
 \node[state, circle split] (n12) [below of = n02]{\_\nodepart{lower}4};  
 \node[state, circle split] (n22) [below of = n12]{\_\nodepart{lower}4};  
 \node[state, circle split] (n32) [below of = n22]{\_\nodepart{lower}1};

\path[->]	(n00) edge [bend left = 10] node{} (n01) 	
		(n01) edge [bend left = 10] node{} (n02) 
		(n02) edge [bend left = 10] node{} (n03) 
		(n00) edge node {} (n10)
		(n10) edge [loop below] node{} ()
		(n01) edge node {} (n11)
		(n11) edge node {} (n21)	
		(n21) edge [loop below] node{} ()
		(n02) edge node {} (n12)
		(n12) edge node {} (n22)	
		(n22) edge node {} (n32)	
		(n32) edge [loop below] node{} ();
 \end{tikzpicture}
 \end{minipage}
 \begin{minipage}{2.5in}
 \begin{tikzpicture}[shorten >=.5pt,node distance=1.9cm, auto]
 \node[state, circle split] (n00){0\nodepart{lower}3};
 \node[state, circle split] (n01) [right of = n00]{0\nodepart{lower}3};
 \node[state, circle split] (n02) [right of = n01]{0\nodepart{lower}3};
 \node (n03) [right of = n02]{$\dots$};   
 \node[state, circle split] (n10) [below of = n00]{\_\nodepart{lower}1};  
 \node[state, circle split] (n11) [below of = n01]{\_\nodepart{lower}4};  
 \node[state, circle split] (n21) [below of = n11]{\_\nodepart{lower}1};  
 \node[state, circle split] (n12) [below of = n02]{\_\nodepart{lower}4};  
 \node[state, circle split] (n22) [below of = n12]{\_\nodepart{lower}4};  
 \node[state, circle split] (n32) [below of = n22]{\_\nodepart{lower}1};

\path[->]	(n00) edge [bend left = 10] node{} (n01) 	
		(n01) edge [bend left = 10] node{} (n02) 
		(n02) edge [bend left = 10] node{} (n03) 
		(n00) edge node {} (n10)
		(n10) edge [loop below] node{} ()
		(n01) edge node {} (n11)
		(n11) edge [loop below] node {} ()	
		(n21) edge [loop below] node{} ()
		(n02) edge node {} (n12)
		(n12) edge  [loop below]node {} ()	
		(n22) edge [loop below] node {} ()	
		(n32) edge [loop below] node{} ();
 \end{tikzpicture}
 \end{minipage}
\caption{$G$ (lefthand side) induces $G^+$ which may need to be adjusted a transfinite number of times.}
\label{fig:transfinite}
 \end{figure}
 
The transfinite modification of $G^+$ is now formally defined by mutual induction with the number of relevant properties. Let us assume that for some ordinal number $\alpha$ there exist sequences $(\pi_\beta)_{\beta < \alpha}$ (priority functions) and $(\tau_\beta)_{\beta < \alpha}$ (\pO\/ strategies in $G_\beta$) satisfying the following, where $G_{\beta} := \langle  V^+,E^+,\pi_\beta\rangle$. For all $\gamma \leq \beta < \alpha$
\begin{enumerate}
\item\label{enum1} A run that is winning for \pO\/ in $G_\gamma$ is also winning in $G_\beta$, and $W^m_1(G_{\gamma}) \subseteq W^m_1(G_{\beta})$. 
\item\label{enum2} $\tau_\beta\mid_{W^m_1(G_{\gamma})} = \tau_\gamma\mid_{W^m_1(G_{\gamma})}$ %$\tau_\beta\mid_{V^+ \setminus (W^m_1(G_{\beta}) \setminus W^m_1(G_{\gamma}))} = \tau_\gamma\mid_{V^+ \setminus (W^m_1(G_{\beta}) \setminus W^m_1(G_{\gamma}))}$. Especially, $\tau_\beta\mid_{W^m_1(G_{\gamma})} = \tau_\gamma\mid_{W^m_1(G_{\gamma})}$.
\item\label{enum3} $\tau_{\beta}$ is optimal for  \pO\/ in $G_{\beta}$.
\item\label{enum4} $f \circ \tau_{\beta}$ makes \pO\/ win from $W^m_1(G_\beta) \setminus \tilde{D}$ in $G$.
\end{enumerate}
The intermediate goal is to define $\pi_\alpha$ (and thus $G_\alpha := \langle  V^+,E^+,\pi_\alpha\rangle $) and $\tau_\alpha$, a \pO\/ strategies in $G_\alpha$, and to show that the above four properties also hold for the extended sequences $(G_\beta)_{\beta < \alpha+1}$ and $(\tau_\beta)_{\beta < \alpha+1}$. 

Let $X_\alpha := \cup_{\beta < \alpha}W^m_1(G_\beta)$,  let $\pi_{\alpha}(\tilde{v}) := k+1$ for all $v \in D \cap  X_\alpha$, and let $\pi_{\alpha}(u) :=\pi^+(u)$ for all $u \in V^+ \setminus (D \cap  X_\alpha)$. Note that $X_0 = \emptyset$ and $G_0 = G^+$. Let $\tau^-_\alpha$ be a \pO\/ strategy such that $\tau^-_\alpha(v) = \tau_\beta(v)$ for all $v \in X_\alpha$, where $\beta < \alpha$ is the least such that $v \in W^m_1(G_{\beta})$. By I.H. on the relevant priorities, let $\tau^+_\alpha$ be a \pO\/ optimal strategy in $G_\alpha$, and let $\tau_\alpha$ be a \pO\/ strategy, also in $G_\alpha$, that coincides  with $\tau^-_\alpha$ on $X_\alpha$ and with $\tau^+_{\alpha}$ on $W^m_1(G_{\alpha}) \setminus X_\alpha$. Let us now show that the above four properties also hold for the extended sequences $(G_\beta)_{\beta < \alpha+1}$ and $(\tau_\beta)_{\beta < \alpha+1}$. 
\begin{enumerate}
\item For all $v \in V^+$ and $\beta < \alpha$, if $\pi_\alpha(v) \neq \pi_\beta(v)$ then $\pi_\alpha(v) = k+1$ and $\pi_\beta(v) = k$, where $k$ is even by assumption. It is then straightforward to show that a run that is winning for \pO\/ in $G_\beta$ is also winning in $G_\alpha$, and subsequently that $W^m_1(G_{\beta}) \subseteq W^m_1(G_{\alpha})$.

\item Let $\gamma < \alpha$ and $v \in W^m_1(G_\gamma)$. Since $W^m_1(G_\gamma) \subseteq X_\alpha$, by definition of $\tau_\alpha$ and $\tau^-_\alpha$ we have $\tau_\alpha(v) = \tau^-_\alpha(v) = \tau_\beta(v)$, where $\beta$ is the least such that $v \in W^m_1(G_{\beta})$. This implies $\beta \leq \gamma$ since $v \in W^m_1(G_\gamma)$ by choice of $v$, so $\tau_\beta\mid_{W^m_1(G_{\beta})} = \tau_\gamma\mid_{W^m_1(G_{\beta})}$ by I.H (item ~\ref{enum2}). Evaluating this equation at $v \in W^m_1(G_{\beta})$ yields $\tau_\beta(v) = \tau_\gamma(v)$, i.e. $\tau_\alpha(v) = \tau_\gamma(v)$. This shows that $\tau_\gamma\mid_{W^m_1(G_{\gamma})} = \tau_\alpha\mid_{W^m_1(G_{\gamma})}$.

\item Let a run $\rho$ in $G_\alpha$ start from $W^m_1(G_\alpha)$ and be compatible with $\tau_\alpha$, and let us make a case disjunction to show that $\tau_\alpha$ is optimal in $G_\alpha$. 
\begin{itemize}
\item First case, $\rho_0 \in X_\alpha = \cup_{\beta < \alpha}W^m_1(G_\beta)$. Let $\beta$ be any ordinal such that $\rho_0 \in W^m_1(G_\beta)$, and let us prove that  $\rho$ is compatible with $\tau_\beta$. If $\rho_0 \in V_0$, then  $\rho_0 \rho_1$ is compatible with $\tau_\beta$, and $\rho_1 \in W^m_1(G_\beta)$; if $\rho_0 \in V_1$, the definition of $\tau_\alpha$ implies that $\rho_1 = \tau_{\alpha}(\rho_0) = \tau_{\gamma}(\rho_0)$, where $\gamma$ is the least such that $\rho_0 \in W^m_1(G_\gamma)$. So $\rho_1 = \tau_{\beta}(\rho_0)$ since $\tau_\beta\mid_{W^m_1(G_{\gamma})} = \tau_\gamma\mid_{W^m_1(G_{\gamma})}$ by I.H (item~\ref{enum2}), so on the one hand $\rho_0 \rho_1$ is compatible with $\tau_\beta$, and on the other hand $\rho_1 \in W^m_1(G_\gamma) \subseteq W^m_1(G_\beta)$: the membership holds since $\tau_{\gamma}$ is optimal for  \pO\/ in $G_{\gamma}$ by I.H. (item~\ref{enum3}), and the inclusion holds by I.H. (item~\ref{enum1}). Invoking these two facts recursively shows that $\rho$ is compatible with $\tau_\beta$. So $\rho$ is winning for \pO\/ in $G_\beta$ by  I.H. (item~\ref{enum3}), and also in $G_\alpha$ by the above item~\ref{enum1}.

% by s also in and that $\rho$ is winning for \pO\/ in $G_\alpha$ since  $\tau_{\beta}$ is optimal for  \pO\/ in $G_{\beta}$ 

\item Second case, $\rho$ enters $X_\alpha$ at some point, so the first case applies to its tail and prefix independence implies that $\rho$ is winning for \pO\/. 

\item Third case, $\rho$ avoids $X_\alpha$, so it stays in $W^m_1(G_\alpha) \setminus X_\alpha$ and is compatible with $\tau^+_\alpha$, so it is also winning for \pO\/.
\end{itemize}

\item Let a run $\rho$ in $G$ start from $W^m_1(G_{\alpha})\setminus \tilde{D}$ and be compatible with $f \circ \tau_\alpha$, and let us make a case disjunction. If $\rho$ never visits $D$ (but possibly at the start), it is also compatible with $\tau_\alpha$ by the equivalence (\ref{eq:comp-run}), and it is winning for \pO\/ since $\tau_\alpha$ is optimal in $G_\alpha$ by the above item~\ref{enum3}. If $\rho$ visits $D$ for the first time at some time $n +1$, then $\tau_\alpha$ and $f \circ \tau_\alpha$ coincide along $\rho_{<n}$ until $\rho_{n+1} \in D$, also by the equivalence (\ref{eq:comp-run}). Since $\rho_{\leq n+1}$ is compatible with $f \circ \tau_\alpha$, also $\rho_{\leq n}\tilde{\rho}_{n+1}$ is compatible with $\tau_\alpha$. So $\pi_{\alpha}(\tilde{\rho}_{n+1}) = k+1$ (instead of $k$) since $\rho_0 \in W^m_1(G_{\alpha})$ by assumption, and since $\tau_\alpha$ is optimal for \pO\/ in $G_\alpha$ by the above item~\ref{enum3}. By construction of $\pi_\alpha$ this implies that $\rho_{n+1} \in W^m_1(G_{\beta})$ for some $\beta < \alpha$. Since $\tau_\alpha$ and $\tau_\beta$ coincide on $W^m_1(G_{\beta})$ by the above item~\ref{enum2}, so do  $f \circ \tau_\alpha$ and $f  \circ \tau_\beta$ (especially on $W^m_1(G_{\beta})\setminus \tilde{D}$ ). Since  $f \circ \tau_{\beta}$ wins from $W^m_1(G_\beta) \setminus \tilde{D}$ in $G$ by I.H. (item~\ref{enum4}), since $\rho_{n+1} \in W^m_1(G_\beta) \setminus \tilde{D}$, and since only absorbing states have self-loops by assumption, a straightforward induction shows that $\rho$ follows $f \circ \tau_{\beta}$ (in addition to $f \circ \tau_\alpha$) and remains in $W^m_1(G_{\beta})\setminus \tilde{D}$ from $\rho_{n+1}$ on. So the corresponding tail of $\rho$ is winning for \pO\/ by I.H. (item~\ref{enum4}). By prefix independence $\rho$ is also winning.
\end{enumerate}

By definition $\beta < \alpha$ implies $X_\beta \subseteq X_\alpha$, so for cardinality reasons there exists a (least) ordinal $\alpha_0$ such that $X_{\alpha_0 +1} = X_{\alpha_0}$. This implies $W^m_1(G_{\alpha_0}) = X_{\alpha_0}$, so 
\begin{align}
\forall v \in D,\quad\pi_{\alpha_0}(\tilde{v}) = k+1\,\Leftrightarrow\,v \in W^m_1(G_{\alpha_0})\label{eq:prior}
\end{align} 
{\bf First witness strategy} By item~\ref{enum4} above, $f \circ \tau_{\alpha_0}$ makes \pO\/ win from $W^m_1(G_{\alpha_0}) \setminus \tilde{D}$ in $G$.

{\bf Second witness strategy} By I.H. (on the relevant priorities) let $\sigma$ be such that $W^m_0(G_{\alpha_0},\sigma) = W^m_0(G_{\alpha_0}) = V^+ \setminus W^m_1(G_{\alpha_0})$. Let us argue that $f \circ \sigma$ wins from $V \setminus W^m_1(G_{\alpha_0})$  in $G$, which will prove the claim since $(V \setminus W^m_1(G_{\alpha_0})) \cup (W^m_1(G_{\alpha_0}) \setminus \tilde{D}) = V$. 

Let a run $\rho$ start from $V \setminus W^m_1(G_{\alpha_0}) = V \cap W^m_0(G_{\alpha_0})$ and be compatible with $f \circ \sigma$, and let us make a case disjunction: if $\rho$ never sees the priority $k$, it avoids $D$ and is also compatible with $\sigma$ by the equivalence~(\ref{eq:comp-run}), and it is winning for \pZ\/ since $\rho_0 \in W^m_0(G_{\alpha_0})$; if $\rho$ sees the priority $k$ infinitely often, it is winning for \pZ\/. 

The remaining case is that $\rho$ sees the priority $k$ first at some time $n+1$, and then only finitely many times. In particular, $\rho_{n+1}$ is not an absorbing vertex. Since it is not a vanishing vertex either, $\rho_{n+1} \in D$. By Equivalence~(\ref{eq:comp-run}), $\sigma$ and $f \circ \sigma$ coincide along $\rho_{<n}$ since priority $k$ (i.e on a vertex in $D$) is first seen at $\rho_{n+1}$. Since $\rho_0 \in W^m_0(G_{\alpha_0})$ and $\sigma$ is optimal for \pZ\/ in $G_{\alpha_0}$, $\pi_{\alpha_0}(\tilde{\rho}_{n+1}) = k$ (as opposed to $k+1$), so  $\rho_{n+1} \in W^m_0(G_{\alpha_0})$ by Equivalence~(\ref{eq:prior}). By applying the argument iteratively, one shows that $\rho$ stays in $W^m_0(G_{\alpha_0})$, and after the last time $\rho$ sees $k$, its tail is winning for \pZ\/ by the first case above.
\end{proof}
\end{theorem}

\bibliographystyle{plain}
\bibliography{article}

\end{document}